\documentclass[jap,twocolumn,groupedaddress,nofootinbib]{revtex4-1}
\usepackage{graphicx}
\usepackage{dcolumn}
\usepackage{bm}

\newcommand{\fepb}{Fe$_5$PB$_2$}
\newcommand{\fesib}{Fe$_5$SiB$_2$}
\newcommand{\fecopb}{Fe$_4$CoPB$_2$}
\newcommand{\fecosib}{Fe$_4$CoSiB$_2$}
\newcommand{\femnpb}{Fe$_4$MnPB$_2$}
\newcommand{\femnsib}{Fe$_4$MnSiB$_2$}

\begin{document}

\title{Magnetic and structural properties of ferromagnetic \fepb\ and \fesib\ and effects of Co and Mn substitutions }

\author{Michael A. McGuire}
\email{McGuireMA@ornl.gov \\  \\ Notice: This manuscript has been authored by UT-Battelle, LLC under Contract No. DE-AC05-00OR22725 with the U.S. Department of Energy. The United States Government retains and the publisher, by accepting the article for publication, acknowledges that the United States Government retains a non-exclusive, paid-up, irrevocable, world-wide license to publish or reproduce the published form of this manuscript, or allow others to do so, for United States Government purposes. The Department of Energy will provide public access to these results of federally sponsored research in accordance with the DOE Public Access Plan(http://energy.gov/downloads/doe-public-access-plan). }
\affiliation{Oak Ridge National Laboratory, Oak Ridge, Tennessee 37831 USA}

\author{David S. Parker}
\affiliation{Oak Ridge National Laboratory, Oak Ridge, Tennessee 37831 USA}

\begin{abstract}

Crystallographic and magnetic properties of \fepb, \fecopb, \femnpb, \fesib, \fecosib, and \femnsib\ are reported. All adopt the tetragonal Cr$_5$B$_3$ structure-type and are ferromagnetic at room temperature with easy axis of magnetization along the \textit{c}-axis. The spin reorientation in \fesib\ is observed as an anomaly in the magnetization near 170 K, and is suppressed by substitution of Co or Mn for Fe. The silicides are found to generally have larger magnetic moments than the phosphides, but the data suggests smaller magnetic anisotropy in the silicides. Cobalt substitution reduces the Curie temperatures by more than 100 K and ordered magnetic moments by 16-20\%, while manganese substitution has a much smaller effect. This suggests Mn moments align ferromagnetically with the Fe and that Co does not have an ordered moment in these structures. Anisotropic thermal expansion is observed in \fepb\ and \fesib, with negative thermal expansion seen along the \textit{c}-axis of \fesib. First principles calculations of the magnetic properties of \fesib\ and \femnsib\ are reported. The results, including the magnetic moment and anisotropy, and are in good agreement with experiment.

\end{abstract}

\maketitle

\section{Introduction}

Uniaxial ferromagnets composed of earth-abundant elements are of interest to researchers exploring new materials for permanent magnet applications \cite{Coey-2011}. \fepb\ and \fesib\ are examples of such materials that have not been heavily investigated in this respect to date. The crystal structures and compositions of these compounds were first investigated by Aronsson and coworkers \cite{Aronsson-1959, Aronsson-1960} and Rundqvist \cite{Rundqvist-1962}. They were shown to be isostructural, adopting the tetragonal Cr$_5$B$_3$ structure type. Ferromagnetism was first reported in the phosphide based on magnetization measurements \cite{Blanc-1967} and M\"{o}ssbauer spectroscopy \cite{Haggstrom-1975}, which identified the uniaxial nature of the magnetic structure with the easy axis of magnetization along the crystallographic \textit{c}-axis. Soon thereafter, \fesib\ was identified as a uniaxial ferromagnet at room temperature as well \cite{Wappling-1976}, and a spin reorientation was seen to occur near 140 K, resulting in planar magnetic anisotropy at lower temperature \cite{Ericsson-1978}. The Curie temperatures ($T_C$) of both phases were found to be well above room temperature, near 630 K for \fepb\ \cite{Blanc-1967} and 784 K for \fesib \cite{Wappling-1976}. A very recent study of \fepb\ reports $T_C$ = 655 K, magnetic anisotropy comparable to hard ferrites, and a magnetic moment per Fe of 1.72 $\mu_B$ \cite{Lamichhane-2015}. The isostructural Mn compounds are also ferromagnetic, but with Curie temperatures closer to room temperature, making them perhaps more interesting as magnetocaloric rather than magnet materials \cite{de-Almeida-2009, Xie-2010}. No reports of magnetization measurements for \fesib\ were located in the literature, and no detailed crystallographic study has been published to date. In addition, understanding of effects of chemical substitutions in these uniaxial, high-temperature ferromagnets is presently lacking.

In this work, we report structural and magnetic properties determined from measurements on polycrystalline samples of \fepb, \fesib, \fecopb, \fecosib, \femnpb, and \femnsib. Full crystal structure refinements for all compounds are reported at room temperature, and for both \fepb\ and \fesib\ at 20 K. Significant anisotropy is observed in the thermal expansion, which is negative along the \textit{c}-direction in \fesib. Magnetization measurements are used to determine the saturation magnetizations and identify the Curie and spin-reorientation temperatures. Results of density functional theory calculation of the ordered magnetic moment and anisotropy of \fesib\ and \femnsib\ are reported, and compared to the measurements. The silicides are found to have a larger magnetic moment than the phosphides, but smaller magnetic anisotropy at 300 K is indicated in the silicides. Both chemical substitutions suppress the spin reorientation in \fesib, indicating \fecosib\ and \femnsib\ are uniaxial ferromagnets down to at least 30 K. In both \fepb\ and \fesib, substitution of Mn expands the unit cell, has little effect on the Curie temperature, and produces a slight reduction in magnetic moment. Substitution of Co contracts the unit cell, and strongly suppresses both the Curie temperature and the magnetic moment. These findings suggest that the ordered magnetic moment on Co in these structures is either very small or aligned non-collinearly with the net magnetization.

\section{Experimental Procedures}

Polycrystalline samples of nominal compositions listed in Table \ref{tab:properties} were synthesized by reacting the elements in evacuated and flame-sealed silica tubes. The tubes were heated to 1000 $^{\circ}$C at 100\,$^{\circ}$C per hour, and held at this temperature for 12 hours. The resulting mixtures were then ground into powders and pressed into 1/2 inch diameter pellets at room temperature. The pellets were sealed in evacuated silica tubes and heated at 1000\,$^{\circ}$C for one week. The pellets were then ground into powder, re-pelletized, and heated in evacuated silica tubes at 1000\,$^{\circ}$C for 5 days. Samples produced in this way were found to be $\gtrsim$90\% pure. For the silicon-containing samples (with no phosphorus), the initial mixing of the elements can be done by arc-melting, but subsequent grinding, pelletizing, and annealing are needed to improve the phase purity. This later route was used to make the Fe$_4$CoSiB$_2$ used in this study, because it gave a final sample with less ferromagnetic impurity.

X-ray diffraction was performed on a PANalytical X-Pert Pro diffractometer using monochromatic Cu K$\alpha_1$ radiation in Bragg-Brentano reflection geometry. An Oxford PheniX cryostat was used for diffraction measurements below room temperature. Quantitative analysis and crystal structure refinement was done with Rietveld refinement using the program FullProf \cite{Fullprof}. To determine the easy axis of magnetization by diffraction, finely ground samples were sprinkled onto a vacuum grease coated glass slide with a strong permanent magnet underneath. Compositions of the primary phases were examined using energy dispersive analysis with a Bruker Quantax 70 x-ray spectrometer (EDS) in a Hitachi TM3000 tabletop scanning electron microscope. Magnetization measurements on solid polycrystalline samples cut from the sintered pellets were performed with a Quantum Design Magnetic Property Measurement System.

\section{Results and Discussion}
\subsection{X-ray diffraction}

\begin{figure*}
\begin{center}
\includegraphics[width=6in]{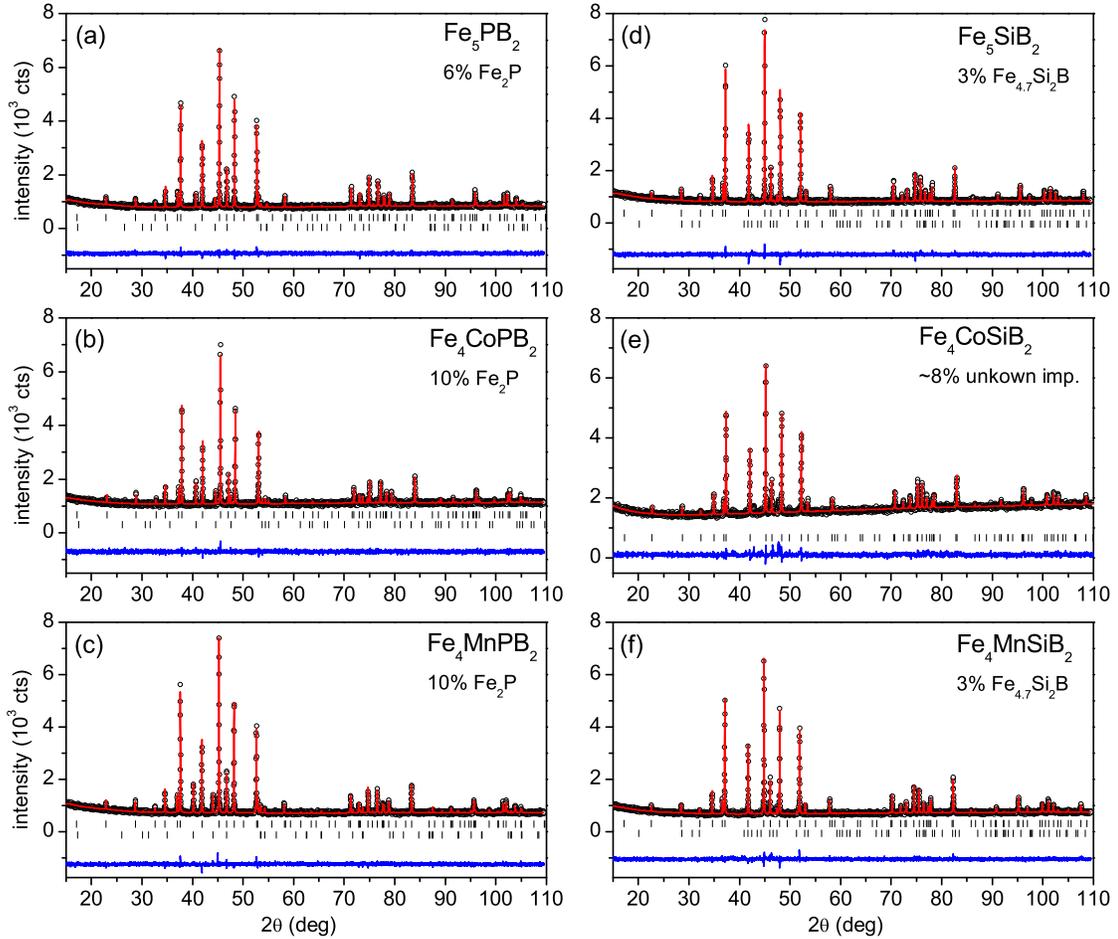}
\caption{\label{fig:pxrd}
Room temperature powder x-ray diffraction patterns and fits from Rietveld refinements. Difference curves are shown at the bottom of each panel. The lower set of tics in panels a-d and f locate Bragg reflections from the impurity phases noted on each plot. The impurity in Fe$_4$CoSiB$_2$ has not been identified and its concentration is estimated from relative peak intensities.
}
\end{center}
\end{figure*}

Analysis of powder x-ray diffraction patterns (Fig. \ref{fig:pxrd}) shows that all of the materials studied adopt the tetragonal Cr$_5$B$_3$ structure type at room temperature. The samples used in the current study were $\gtrsim$90\% pure, and the impurities identified in the diffraction patterns along with their weight percent concentrations are shown on the Figure. In the Fe$_5$PB$_2$-based materials, the only secondary phase detected was Fe$_2$P, which is paramagnetic at room temperature, and undergoes a first-order ferromagnetic phase transition near 216 K \cite{Lundgren-1978}. About 3 \% of Fe$_{5-\delta}$Si$_2$B \cite{Aronsson-1959, Aronsson-1960} was observed in the Fe$_5$SiB$_2$ and Fe$_4$MnSiB$_2$ samples. The magnetic properties of this phase have not been reported, but in the course of this study we have synthesized and performed preliminary characterization of a sample of composition Fe$_{4.7}$Si$_2$B and find that it is likely paramagnetic at room temperature and antiferromagnetic at lower temperature. The Fe$_4$CoSiB$_2$ sample (Fig. \ref{fig:pxrd}e) contains a small amount of an unidentified impurity phase, which may be reflected in the high temperature magnetization data presented below.

\begin{table*}
\begin{center}
\caption{\label{tab:properties} Crystallographic and magnetic properties. All materials adopt the Cr$_5$B$_3$ structure type with space group $I4/mcm$ and $M$1 at 16$l$ (x, x+1/2, z), $M$2 at 4$c$ (0, 0, 0), P/Si at 4$a$ (0, 0, 1/4), B at 8$h$ (x, 1/2+x, 0). Note $J = \mu_0M$, and divide $\textit{J}$ in Tesla by $4\pi\times 10^{-4}$ to convert to \textit{M} in emu/cm$^3$, or by $4\pi\times 10^{-7}$ to convert to \textit{M} in A/m.
}
\begin{tabular}{lrrrrrr}
\hline													
nominal composition	&	Fe$_5$PB$_2$	&	Fe$_4$CoPB$_2$	&	Fe$_4$MnPB$_2$	&	Fe$_5$SiB$_2$	&	Fe$_4$CoSiB$_2$	&	 Fe$_4$MnSiB$_2$	\\
EDS (Co,Mn)/Fe ratio	&	--	&	0.28	&	0.24	&	--	&	0.25	&	0.26	\\
\hline
\textit{a} ({\AA})	&	5.4867(1)	&	5.4522(1)	&	5.4947(1)	&	5.5507(1)	&	5.5332(2)	&	5.5689(1)	\\
\textit{c} ({\AA})	&	10.3532(2)	&	10.3585(3)	&	10.3754(2)	&	10.3359(2)	&	10.2692(4)	&	10.3582(2)	\\
\textit{V} ({\AA}$^3$)	&	311.67(1)	&	307.92(1)	&	313.25(1)	&	318.45(1)	&	314.40(2)	&	321.24(1)	 \\
density (g / cm$^3$)	&	7.075	&	7.227	&	7.020	&	6.864	&	7.017	&	6.785	\\
\textit{c/a}	&	1.887	&	1.900	&	1.888	&	1.862	&	1.856	&	1.860	\\
$x_{M1}$	&	0.1701(4)	&	0.1698(5)	&	0.1699(4)	&	0.1678(4)	&	0.1694(5)	&	0.1685(4)	\\
$z_{M1}$	&	0.1400(3)	&	0.1419(4)	&	0.1410(3)	&	0.1384(3)	&	0.1387(4)	&	0.1381(3)	\\
$x_B$	&	0.384(4)	&	0.385(5)	&	0.388(4)	&	0.386(4)	&	0.372(5)	&	0.382(4)	\\
$\chi^2$	&	1.2	&	1.3	&	1.5	&	1.3	&	1.2	&	1.4	\\
\hline													
$J_{S}$ 5 K (T) &	            1.20 	&	1.03 	&	1.18 	&	1.31 	&	1.06 	&	1.22 	\\
$M_{S}$ 5 K ($\mu_B$/F.U.)	&	8.00	&	6.71	&	7.51	&	9.06	&	7.24	&	8.52	\\
$J_{S}$ 300 K (T) 	&	1.09 &	0.87 &	1.00 &	1.21 &	0.961 &	1.17 \\
$M_{S}$ 300 K ($\mu_B$/F.U.)	&	7.29	&	5.68	&	6.72	&	8.39	&	6.55	&	8.14	\\
$T_C$ (K)	&	640	&	515	&	650	&	$\gtrsim$ 780	&	675	&	770	\\

\hline													
\end{tabular}
\end{center}
\end{table*}

\begin{figure}
\begin{center}
\includegraphics[width=3.5in]{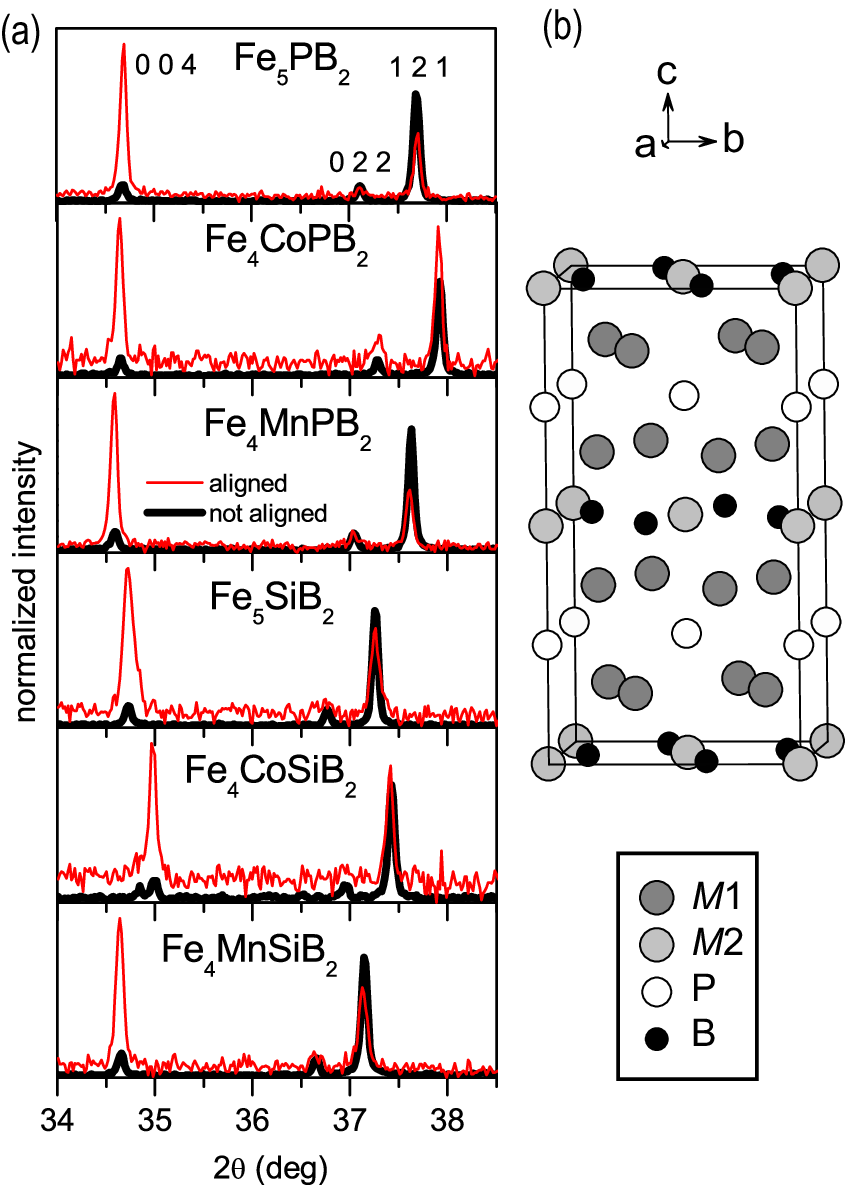}
\caption{\label{fig:aligned-pxrd}
(a) Powder x-ray diffraction results showing the results of alignment of the powders with a magnetic field normal to the sample surface. (b) The crystal structure adopted by the materials studied here.
}
\end{center}
\end{figure}

Crystallographic data determined from refinement of the diffraction data are collected in Table \ref{tab:properties}. The structure determined for Fe$_5$PB$_2$ and listed in Table \ref{tab:properties} is in good agreement with that originally reported by Rundqvist and coworkers \cite{Rundqvist-1962, Haggstrom-1975} and confirmed recently with single crystal x-ray diffraction \cite{Lamichhane-2015}. The unit cell parameters of Fe$_5$SiB$_2$ in Table \ref{tab:properties} are also consistent with literature reports \cite{Aronsson-1960}, although a full structure refinement has not been previously reported. The crystal structure is shown in Fig. \ref{fig:aligned-pxrd}b. Sites labeled \textit{M1} and \textit{M2} are occupied by the transition metal atoms. At \textit{z} = 0 and $\frac{1}{2}$ are planes of composition \textit{M2}B$_2$. Slabs of composition \textit{M1}$_4$P fill the space between these planes. The \textit{M1} positions has site symmetry \textit{m} and an irregular coordination environment. It is coordinated by two P atoms, three B atoms, two \textit{M}2 sites, and two \textit{M}1 sites.  \textit{M}2 with site symmetry 4/\textit{m}  is at the center of a square of four B atoms in the \textit{ab}-plane and a slightly distorted cube of eight \textit{M}1 atoms.

The ratio of transition metals in the main phase of the Co and Mn substituted samples was determined by EDS and is shown in Table \ref{tab:properties}. This was determined by averaging multiple measurements on single grains of the primary phase. The results are consistent with the nominal ratio of 1:4. Allowing mixing of P or Si and B on their respective 4\textit{a} and 8\textit{h} sites did not result in any improvement of the fits to the diffraction data for any of the samples studied here, but it did indicate that some B may replace the heavier main-group elements, as reported previously in \fepb\ \cite{Haggstrom-1975, Lamichhane-2015}. For the final structural refinements occupancies were fixed at the nominal composition.

Substitution of Co is seen to reduce the lattice parameters and unit cell volume, while substitution of Mn has the opposite effect. The $c/a$ ratio is not substantially affected by the substitutions, but it is slightly different between the P- and Si-containing materials. These changes are as expected based on the published lattice constants of Co$_5$PB$_2$ (5.42 {\AA}, 10.20 {\AA}) \cite{Rundqvist-1962}, Mn$_5$PB$_2$ (5.54 {\AA}, 10.49 {\AA}) \cite{Rundqvist-1962}, and Mn$_5$SiB$_2$ (5.61 {\AA}, 10.44 {\AA}) \cite{Aronsson-1959}. Note that Co$_5$SiB$_2$ is not known to form.

All of the materials are strongly ferromagnetic at room temperature (\textit{vide infra}). M\"{o}ssbauer spectroscopy has shown that Fe$_5$PB$_2$ and Fe$_5$SiB$_2$ are uniaxial ferromagnets at room temperature, with the easy axis of magnetization along the tetragonal \textit{c}-axis \cite{Haggstrom-1975, Wappling-1976, Ericsson-1978}. To confirm this in the samples studied here, and to determine if the alloying of Co or Mn may change the easy axis, diffraction patterns were collected from powders aligned in a magnetic field so that the easy axis of magnetization was normal to the sample plane. The results are shown in Fig. \ref{fig:aligned-pxrd}a. Three reflections are observed in the angular range shown. They are labeled by their indices on the top panel. For each material, the intensity of the 004 reflection is strongly increased relative to the 022 and 121 reflections when the powders are magnetically aligned. This indicates that the easy axis of magnetization is parallel to the \textit{c}* reciprocal lattice vector. For tetragonal symmetry, this is the \textit{c}-axis of the crystal structure. Some intensity is still observed on the 022 and 121 reflections of the aligned samples, which is expected if the alignment is incomplete or if all of the powder particles are not single crystals, but instead are composed of agglomerates of misaligned crystallites.

\begin{figure}
\begin{center}
\includegraphics[width=3.5in]{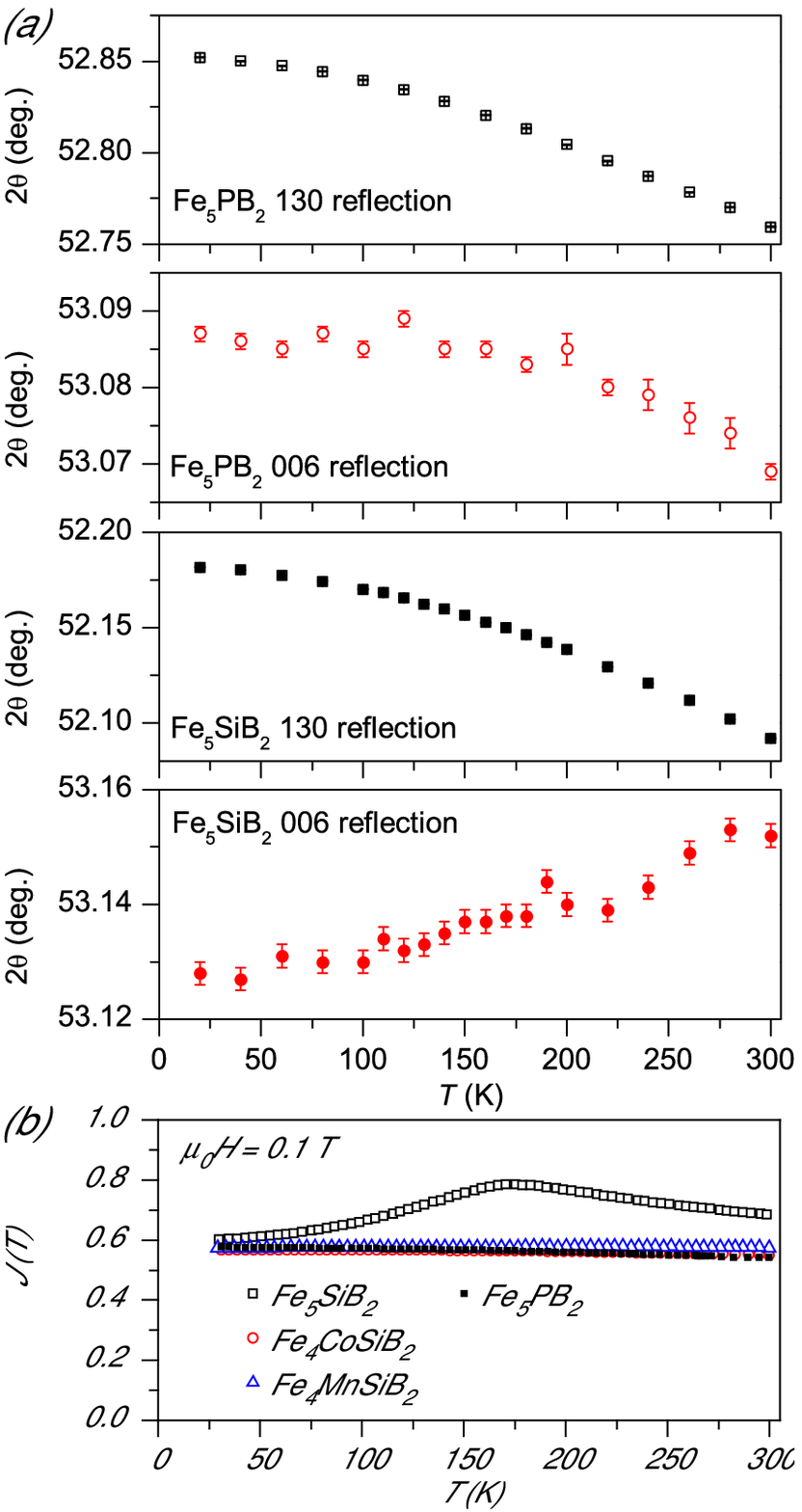}
\caption{\label{fig:thermal-expansion}
Temperature dependence of (a) the positions of the 130 and 006 Bragg reflections from Fe$_5$PB$_2$ and Fe$_5$SiB$_2$ and (b) the magnetization measured approximately along the room temperature easy axis for \fesib, \fecosib, \femnsib, and \fepb.
}
\end{center}
\end{figure}
\begin{table}
\begin{center}
\caption{\label{tab:20K} Crystallographic properties at 20 K in the Cr$_5$B$_3$ structure type with space group $I4/mcm$ and Fe1 at 16$l$ (x, x+1/2, z), Fe2 at 4$c$ (0, 0, 0), P/Si at 4$a$ (0, 0, 1/4), B at 8$h$ (x, 1/2+x, 0).
}
\begin{tabular}{lrr}
\hline													
 	&	\fepb\	&	\fesib\	\\
\hline
\textit{a}	&	5.4766(1)	&	5.4812(1)	\\
\textit{c}	&	10.3488(2)	&	10.3390(2)	\\
\textit{V}	&	310.39(1)	&	317.46(1)	\\
$x_{Fe1}$	&	0.1698(3)	&	0.1676(3)	\\
$z_{Fe1}$	&	0.1407(2)	&	0.1379(2)	\\
$x_B$	&	0.388(2)	&	0.388(3)	\\
$\chi^2$	&	1.8	&	1.5	\\
\hline													
\end{tabular}
\end{center}
\end{table}

Powder x-ray diffraction patterns from Fe$_5$PB$_2$ and Fe$_5$SiB$_2$ were also collected at 20 K. The results from both materials were well described by the room temperature structures, with lattice parameters 5.4766(1) {\AA} and 10.3488(2) {\AA} for the phosphide and 5.5412(1) {\AA} and 10.3390(2) for the silicide. Other crystallographic parameters determined at 20 K are shown in Table \ref{tab:20K}. A crystallographic distortion may be expected to occur at the spin reorientation temperature in \fesib; however, no distortion was resolved in the data, indicating the magnetoelastic coupling in the \textit{ab}-plane is not strong. Comparison of lattice constants at room temperature (Table \ref{tab:properties}) and 20 K (Table \ref{tab:20K}) reveal strongly anisotropic behavior. In Fe$_5$PB$_2$, the contraction of \textit{a} upon cooling from room temperature to 20 K is more than four times larger than the relative contraction along \textit{c}. In Fe$_5$SiB$_2$, the basal plane also changes more strongly with temperature than the $c$-axis. In fact, the $c$-axis shows negative thermal expansion in this compound. The thermal expansion was examined further by monitoring the 130 and 006 Bragg reflections upon cooling from 300 to 20 K. The results are shown in Figure \ref{fig:thermal-expansion}a. The position of the 130 reflection depends only on the basal plane lattice constant \textit{a} and the 006 position depends only on the \textit{c}-axis length. Normal behavior is seen in Fe$_5$PB$_2$. The negative thermal expansion along \textit{c} is apparent in the temperature dependence of the position of the 006 reflection in Fe$_5$SiB$_2$. This compound is known to undergo a spin reorientation below room temperature, with the easy axis reported to be in the \textit{ab}-plane at low temperature \cite{Ericsson-1978}. Results of magnetization measurement with the field applied approximately along the easy axis of magnetization are shown in Figure \ref{fig:aligned-pxrd}b. For these measurements, finely ground powders were mixed in epoxy and cured at room temperature in a magnetic field. The spin reorientation is apparent near 170 K in \fesib. As expected, no similar anomaly is seen in \fepb, which remains uniaxial at low temperatures \cite{Haggstrom-1975, Lamichhane-2015}. In addition, the Mn and Co substitutions studied here appear to suppress the spin reorientation in \fesib. No lattice anomalies are apparent in Figure \ref{fig:aligned-pxrd}a at the spin reorientation temperature in \fesib.

\subsection{Magnetization measurements}

\begin{figure}
\begin{center}
\includegraphics[width=3.25in]{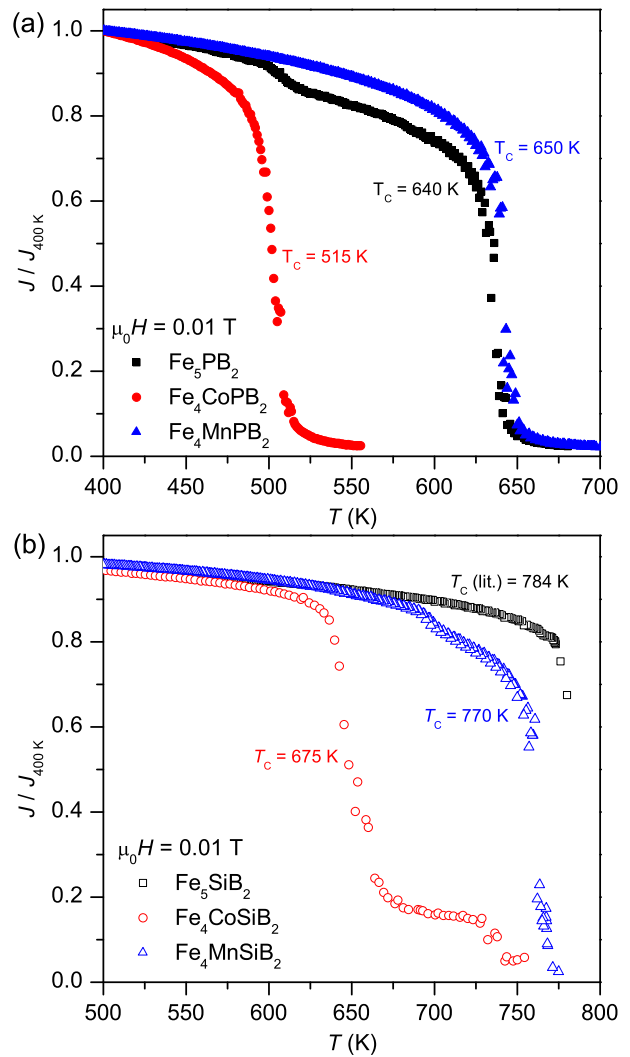}
\caption{\label{fig:JvsT}
Magnetization vs temperature measured in an applied field of 0.01 T for the P-containing compounds (a) and the Si-containing compounds (b). The data were collected on cooling and are normalized to the values measured at 400 K. Curie temperatures estimated from the data are noted on the figure.
}
\end{center}
\end{figure}

High temperature magnetization measurements were conducted to determine the Curie temperature ($T_C$) of each material. The results are shown in Figure \ref{fig:JvsT}. The measurements were performed in a relatively low applied field of 0.01 T to obtain sharp changes in $J(T)$ at $T_C$. The $T_C$ values are listed on Figure \ref{fig:JvsT} and in Table \ref{tab:properties}, and were determined by extrapolating to $J$ = 0 the steepest part of the magnetization curve. Relatively small contributions to the magnetization arising from ferromagnetic impurity phases are seen in Fe$_5$PB$_2$ (Fig. \ref{fig:JvsT}a), Fe$_4$CoSiB$_2$, and Fe$_4$MnSiB$_2$ (Fig. \ref{fig:JvsT}b). The identities of the additional ferromagnetic phases is not clear; however, the magnetic impurity in Fe$_4$CoSiB$_2$ with $T_C$ near 725 K is likely associated with the unidentified reflections in the x-ray diffraction data for this material (Fig. \ref{fig:pxrd}e) based on diffraction and magnetization measurements on samples of varying purity.

The Curie temperature of Fe$_5$SiB$_2$ could not be determined because it is above 780 K, the maximum temperature for these measurements. The downturn seen at the highest temperatures indicates $T_C$ is near but larger than this value. This is consistent with the literature report of $T_C$ = 784 K for Fe$_5$SiB$_2$, although some change in $T_C$ may be expected to occur with variation in Si/B or P/B stoichiometry in these compounds \cite{Haggstrom-1975, Wappling-1976}. Measurements on Fe$_5$PB$_2$ (Fig. \ref{fig:JvsT}a) give $T_C$ = 640 K, similar to the reported values of 615$-$639 K \cite{Blanc-1967} and 655 K \cite{Lamichhane-2015}.

Substitution of 20\% Mn for Fe has a small effect on the determined $T_C$ values. In Fe$_5$PB$_2$ $T_C$ is increased by 10 K, while in Fe$_5$SiB$_2$ it is decreased by 14 K. These variations are within the limits that might be expected for P-B or Si-B antisite defects, based on the the report for Fe$_5$P$_{1-x}$B$_{2+x}$ \cite{Blanc-1967}. A more dramatic change is observed for Co substitution. Replacing one of five Fe atoms with Co reduces $T_C$ by more than 100 K in both Fe$_5$PB$_2$ and Fe$_5$SiB$_2$.

\begin{figure}
\begin{center}
\includegraphics[width=3.25in]{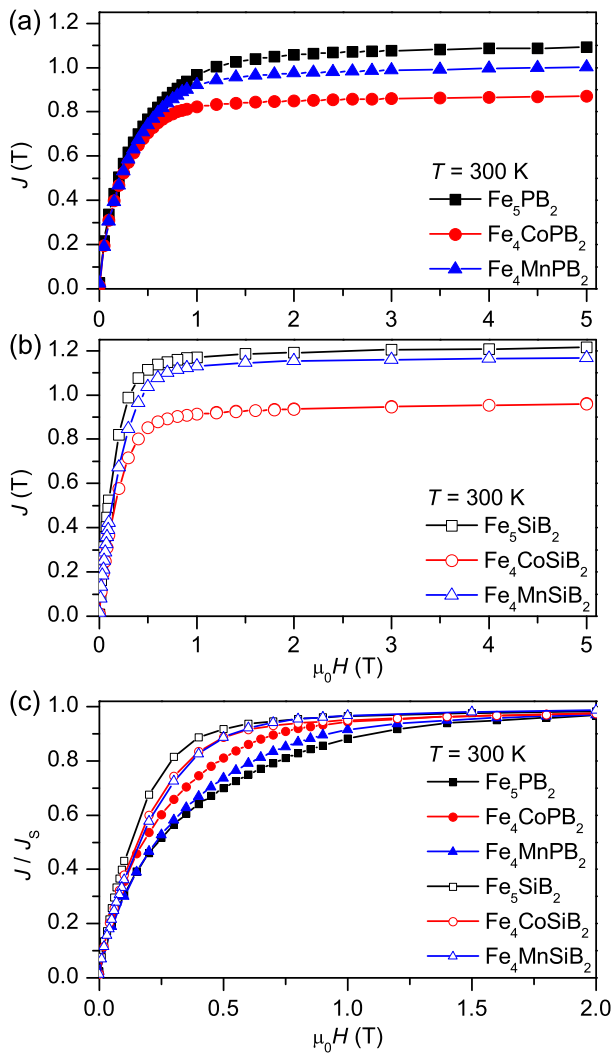}
\caption{\label{fig:JvsH}
Magnetization vs applied magnetic field measured at 300 K for the P-containing compounds (a) and the Si-containing compounds (b). Magnetization normalized by the saturation value is shown in (c) to facilitate comparisons of anisotropy fields. Note $J = \mu_0M$, and divide $\textit{J}$ in Tesla by $4\pi\times 10^{-4}$ to convert to \textit{M} in emu/cm$^3$, or by $4\pi\times 10^{-7}$ to convert to \textit{M} in A/m.
}
\end{center}
\end{figure}

Results of magnetization measurements at 300K are shown in Figure \ref{fig:JvsH}. All materials show ferromagnetic behavior. The saturation magnetization ($J_S$), defined here as the value measured at an applied field of 5 T, for each material is listed in Table \ref{tab:properties}. Determinations of $J_S$ were also performed at 5 K and the results are also listed. A low temperature extrapolation value of 1.73 $\mu_B$/Fe has been reported for the magnetic moment in Fe$_5$PB$_2$ \cite{Blanc-1967}. Recently Lamichhane \textit{et al.} reported the saturation magnetization of this compound to be 1.72 $\mu_B$/Fe at 2 K and about 910 kA/m (1.1 T) at 300 K \cite{Lamichhane-2015}. The present measurements (Table \ref{tab:properties}) are in reasonable agreement with these reports. In general, the magnetization is higher in the Si-based compounds than in the analogous P-based compounds. This is consistent with room temperature M\"{o}ssbauer spectroscopy studies which find larger hyperfine fields at the Fe nuclei in Fe$_5$SiB$_2$ (17$-$23 T) than in Fe$_5$PS$_2$ (15$-$21 T) \cite{Haggstrom-1975, Ericsson-1978}.

Substitution of Co strongly reduces the magnetic moment, as seen in Figure \ref{fig:JvsH} and Table \ref{tab:properties}. Comparing $J_S$ at 5 K, a reduction of 16\% and 20\% are observed when 1/5 of the Fe is replaced by Co in Fe$_5$PB$_2$ and Fe$_5$SiB$_2$, respectively. Assuming that the average magnetic moment on Fe remains the same when Co is substituted gives an average ordered moment per Co of 0.31 $\mu_B$ in Fe$_4$CoPB$_2$ and 0.00 $\mu_B$ in Fe$_4$CoSiB$_2$. The absence of a large ordered moment on Co is consistent with our preliminary measurements, which suggest the absence of ferromagnetism in Co$_5$PB$_2$. Substitution of Mn has a smaller effect on the magnetization. In both Fe$_5$PB$_2$ and Fe$_5$SiB$_2$, substituting 1/5 of the Fe with Mn reduces $J_S$ by 6\% at 5 K. This suggests that Mn does have an ordered moment in these compounds. Indeed, the pure Mn analogues are known to be ferromagnetic \cite{de-Almeida-2009, Xie-2010}. Assuming no change in the average moment per Fe atom occurs with Mn substitution, the data in Table \ref{tab:properties} gives 0.85 $\mu_B$ and 1.27 $\mu_B$ per Mn at 5 K in Fe$_4$MnPB$_2$ and Fe$_4$MnSiB$_2$, respectively. The conclusion that substituted Mn is ferromagnetic while substituted Co is not may explain the sharp decrease in $T_C$ that accompanies the addition of Co but not Mn (Fig. \ref{fig:JvsT} and Table \ref{tab:properties}).

The rate at which the magnetization approaches saturation (Fig. \ref{fig:JvsH}) is determined by the magnetic anisotropy. This can be parameterized by the anisotropy field  $H_A$, which is the applied field required to saturate the material in the magnetically hard direction. Since a polycrystalline material with randomly oriented crystallites contains some fraction with the hard axis along the measurement direction, \textit{J} vs \textit{H} will also saturate near the anisotropy field, but the approach is typically quite gradual and $H_A$ can be difficult to define precisely in this way. In some cases, the singular point detection technique \cite{Asti-1974, Bolzoni-2004} can be used to identify anomalies in d$^2J$/d$H^2$ which occur at $H_A$, but that was not successful using the data for the present samples. This technique requires that there is a sharp (ideally discontinuous) change in the slope of $J$($H$) at $H_A$ in the hard direction, so that the second derivative contains a local minimum at this field. But this minimum occurs on top of a smoothly changing background arising from the portions of the sample with the hard axis aligned in other directions. Better estimates can typically be obtained from measurements on single crystals or magnetically aligned powders. This was recently done for \fepb\ single crystals \cite{Lamichhane-2015}, and $K_1$ was determined to be 0.38 MJ/m$^3$ at room temperature. The anisotropy field can then be estimated by $\mu_0H_A = 2K_1/J_S$ to be 0.83 T. An anisotropy field closer to 1 T might be estimated from the data shown in Fig. \ref{fig:JvsH}c, and this disagreement might be due to the neglect of demagnetization effects in the present analysis. With these limitations in mind, the data in Fig. \ref{fig:JvsH}c suggest that Co and Mn substitution reduces $H_A$ in \fepb, but may slightly increase $H_A$ in \fesib. Measurements on single crystals would be desirable to confirm this.

\subsection{First principles calculations}

First principles calculations of Fe$_5$PB$_2$ and Fe$_5$SiB$_2$ were performed to attain theoretical insight and understanding of the magnetic properties of these materials, using the plane-wave density functional theory code WIEN2K \cite{wien}.  We used the generalized gradient approximation of Perdew, Burke and Ernzerhof \cite{pbe}.  For the P compound we used sphere radii of 1.77, 2.15 and 2.06 Bohr for B, Fe and P and an RK$_{max}$ of 7.0 was used.  Here RK$_{max}$ is the product of the smallest sphere radius and the largest plane wave vector.  For the Si compound we used sphere radii of 1.73, 2.24 and 1.82 Bohr for B, Fe and Si, respectively, with an RK$_{max}$ of 7.0.  Approximately 1000 k-points in the full Brillouin zone were used for these calculations, with spin-orbit coupling included (excepting the internal coordinate optimization).

We briefly summarize the results for the P compound (a more detailed discussion of these calculations can be found in Ref. \onlinecite{Lamichhane-2015}). We find an average saturation moment of 1.79 $\mu_{B}$/Fe, in good agreement with our experimental value; this value includes small negative moments ($\sim -0.05 \mu_B$) on the P and B atoms and a small orbital Fe moment of approximately 0.03 $\mu_{B}$/Fe.  From the calculations we find uniaxial anisotropy, as in the experiment, with a magnetic anisotropy constant K$_{1}$ of 0.46 MJ/m$^{3}$, which yields an anisotropy field H$_{A}$ of approximately 0.9 Tesla, in reasonable agreement with the experiment.

The situation for the Fe$_5$SiB$_2$ compound is significantly more complex and interesting.  As with the P compound we find a ferromagnetic state with a saturation moment  of 1.83 $\mu_{B}$/Fe , including negative moments of -0.06 $\mu_{B}$ on the Si and -0.1 $\mu_{B}$ on the B and a small orbital moment of 0.04 $\mu_{B}$ per Fe. This result compares well with the measured value of 9.06 $\mu_B$ per formula unit or 1.81 $\mu_B$ per Fe(Table \ref{tab:properties}).  We present the calculated density-of-states in Fig. \ref{fig:DOS} below, which depicts a substantial exchange splitting of 2-3 eV as well as an electronic structure dominated around the Fermi level by the Fe, as expected in an Fe-based ferromagnetic material.

\begin{figure}
\includegraphics[width=3.25in]{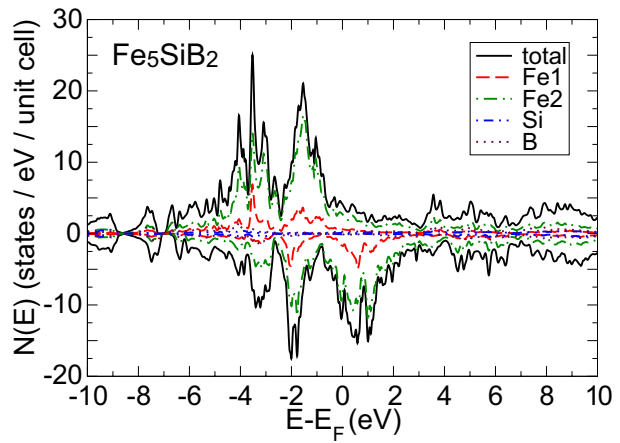}
\caption{\label{fig:DOS}
The calculated density-of-states of Fe$_5$SiB$_2$}
\end{figure}

For the calculations of magnetic anisotropy we used higher precision, with as many as 10,000 k-points in the full Brillouin zone used in the self-consistent runs.  Unlike in the P compound \cite{Lamichhane-2015}, however, we find {\it planar} anisotropy with a first anisotropy constant K$_{1}$ of -0.42 MJ/m$^{3}$. Although \fesib\ is a uniaxial magnet at room temperature, a spin reorientation occurs below room temperature as noted above and first observed in Ref. \onlinecite{Ericsson-1978}, so that our calculation is consistent with the ground state \textit{T} = 0 configuration.  Note that the theoretical calculations were performed using a 300 K structure (with internal coordinates optimized), so that the observed spin-reorientation does not appear to be the result of lattice expansion but may rather originate in dynamical effects. These two scenarios were hypothesized \cite{Shanavas-2014, Antropov-2014} as competing explanations for the spin reorientation \cite{Roberts-1956, Andersen-1967, McGuire-2014} in the manganese-based ferromagnet MnBi, which is planar below 90 K and uniaxial above.  Here we suggest that dynamical effects from the variation in magnetic anisotropy associated with the motion of the ions away from their equilibrium positions are responsible for the spin-reorientation.  This is an effect distinct from both that of lattice expansion as well as that of the general reduction of magnetic moment with increasing temperature, which generally reduces magnetic anisotropy.

We have also performed calculations of the magnetic properties (using the virtual crystal approximation) of the partially Mn substituted compound \femnsib.  As with the other compounds we find a strongly ferromagnetic ordering, with average spin moment 1.88 $\mu_B$ per transition metal, which is slightly larger than the experimental value of 1.50 $\mu_B$ per transition metal (Table \ref{tab:properties}).  It is possible that some of the alloyed Mn anti-align with the Fe, which would reduce the overall moment.  Unlike in the pure compound \fesib\ we find {\it uniaxial} anisotropy in \femnsib.  The calculated anisotropy constant K$_{1}$ is 0.26 MJ/m$^{3}$, yielding an anisotropy field of 0.47 T, in reasonable agreement with the experimental results from Fig. 5.  While a microscopic explanation for the change in anisotropy with Mn alloying is not readily available, we note that similar effects have been calculated to occur in other Fe-containing compounds such as Fe$_{3}$Sn\cite{Sales-2014}. Thus, the present result should not be considered anomalous or unusual, and may motivate further study of Mn substitution as a way to manipulate the magnetic anisotropy in other Fe compounds.

\section{Summary and Conclusions}

We find the tetragonal Cr$_5$B$_3$ structure-type for all compositions and temperatures studied. Anisotropic thermal expansion is observed in \fepb\ and \fesib, with a relative change in \textit{a} being 4$-$6 times larger than that of $c$ between 300 and 20 K. \fesib\ in fact displays negative thermal expansion along the $c$-axis over this temperature range. All of the compounds have the easy axis of magnetization along the crystallographic \textit{c}-axis at room temperature. The saturation moment of \fesib\ is measured to be 1.81 $\mu_B$/Fe at 5 K, in good agreement with density functional theory calculations which find 1.83 $\mu_B$/Fe. Inspection of isothermal magnetization curves at 300 K suggest that \fesib\ has a smaller magnetic anisotropy than \fepb. Calculations (\textit{T} = 0) predict planar anisotropy for \fesib, in agreement with experimental findings \cite{Ericsson-1978}, of similar magnitude but opposite sign from that found in \fepb\ \cite{Lamichhane-2015}. The substitution of Co and Mn does not strongly change the strength of the anisotropy near room temperature, but may slightly decrease it in \fepb\ and slightly increase it in \fesib. An anomaly in the magnetization associated with the rotation of the spins into the \textit{ab}-plane is observed near 170 K in \fesib. No such anomaly is seen in \fepb, \fecosib, or \femnsib\ down to 30 K, showing that the substitution of Co or Mn for Fe suppresses the spin reorientation. Our calculations agree with this finding, showing uniaxial anisotropy in \femnsib. Substitution of Co is seen to strongly suppress the ferromagnetic behavior, reducing the magnetic moment by about 20\% and the Curie temperature by more than 100 K. Mn substitution has a much smaller effect. This is consistent with Co being non-magnetic (which is non-intuitive) and Mn coupling mostly ferromagnetically with the Fe in these structures. The results presented here demonstrate chemical tuning of the strength and anisotropy of the ferromagnetism in \fepb\ and \fesib, and that first principles calculations can reliably predict the magnetic moments and easy axis of magnetization. Therefore we expect that continued work involving closely coupled experiment and theory efforts may identify compositional modifications that would lead to enhanced permanent magnet properties in this family of compounds and others.

\section*{Acknowledgements}
Experimental work (M.A.M.) was supported by the U. S. Department of Energy, Office of Energy Efficiency and Renewable Energy, Vehicle Technologies Office, as part of the Propulsion Materials Program. Theoretical work (D.S.P.) was supported by the Critical Materials Institute, an Energy Innovation Hub funded by the US Department of Energy, Energy Efficiency and Renewable Energy, Advanced Manufacturing Office.


%

\end{document}